\documentclass[aps,reprint,twocolumn,amssymb,superscriptaddress,showpacs]{revtex4-1}
\usepackage[]{fontenc}
\usepackage[latin9]{inputenc}
\usepackage{units}
\usepackage{amstext}
\usepackage{amssymb}
\usepackage{graphicx}

\makeatletter
\@ifundefined{textcolor}{}
{%
 \definecolor{BLACK}{gray}{0}
 \definecolor{WHITE}{gray}{1}
 \definecolor{RED}{rgb}{1,0,0}
 \definecolor{GREEN}{rgb}{0,1,0}
 \definecolor{BLUE}{rgb}{0,0,1}
 \definecolor{CYAN}{cmyk}{1,0,0,0}
 \definecolor{MAGENTA}{cmyk}{0,1,0,0}
 \definecolor{YELLOW}{cmyk}{0,0,1,0}
}

\makeatother

\usepackage[english]{babel}
\usepackage{ragged2e}
\begin{document}

\title{Experimental exploration of the optomechanical attractor diagram and its dynamics}

\author{F.M. Buters}
\email{buters@physics.leidenuniv.nl}
\affiliation{Huygens-Kamerlingh Onnes Laboratorium, Universiteit Leiden,
2333 CA Leiden, The Netherlands}
\author{H.J. Eerkens}
\affiliation{Huygens-Kamerlingh Onnes Laboratorium, Universiteit Leiden,
2333 CA Leiden, The Netherlands}
\author{K. Heeck}
\affiliation{Huygens-Kamerlingh Onnes Laboratorium, Universiteit Leiden,
2333 CA Leiden, The Netherlands}
\author{M.J. Weaver}
\affiliation{Department of Physics, University of California, Santa Barbara,
California 93106, USA}
\author{B. Pepper}
\affiliation{Department of Physics, University of California, Santa Barbara,
California 93106, USA}
\author{S. de Man}
\affiliation{Huygens-Kamerlingh Onnes Laboratorium, Universiteit Leiden,
2333 CA Leiden, The Netherlands}
\author{D. Bouwmeester}
\affiliation{Huygens-Kamerlingh Onnes Laboratorium, Universiteit Leiden,
2333 CA Leiden, The Netherlands}
\affiliation{Department of Physics, University of California, Santa Barbara,
California 93106, USA}

\date{\today{}}

\begin{abstract}
We demonstrate experimental exploration of the attractor diagram of an optomechanical system where the optical forces compensate for the mechanical losses. In this case stable self-induced oscillations occur but only for specific mirror amplitudes and laser detunings. We demonstrate that we can amplify the mechanical mode to an amplitude 500 times larger than the thermal amplitude at 300K. The lack of unstable or chaotic motion allows us to manipulate our system into a non-trivial steady state and explore the dynamics of self-induced oscillations in great detail.

\end{abstract}

\maketitle
\begin{justify}

\section{Introduction}
Laser or microwave cooling of a mechanical degree of freedom has led several groups to come close to or even reach the quantum-mechanical ground state of
a macroscopic harmonic oscillator \cite{teufel2011sideband,chan2011laser,riviere2011optomechanical}.
This has opened up many new research avenues to investigate the foundations
of quantum mechanics \cite{pikovski2012probing}, novel decoherence
mechanisms \cite{kleckner2008creating,pepper2012optomechanical,arndt2014testing}
and strong photon-phonon coupling \cite{groblacher2009observation,nunnenkamp2011single,verhagen2012quantum}.
Besides cooling, also heating of the mechanical degree of freedom
is possible, leading to parametric instabilities, self-induced oscillations
and even chaos. Braginsky et al. have derived the condition for achieving
parametric instability in a Fabry-Perot interferometer such as LIGO \cite{braginsky2001parametric}, 
which is still a topic of interest \cite{cohadon2014parametric}. The theoretical
framework has been expanded by Marquardt et al., with the introduction
of an attractor diagram and an expression for the optomechanical gain
\cite{marquardt2006dynamical,ludwig2008optomechanical}.
From an experimental point of view Carmon et al. showed how self-induced
oscillations of the mechanical mode are imprinted on the cavity output
field \cite{carmon2005temporal,kippenberg2005analysis}.
Finally the transition from self-induced oscillation to chaos has
been investigated with some interesting prospects for observing the
quantum to classical transition \cite{marino2013coexisting,bakemeier2014route}.

The dynamics of self-induced oscillations are best understood using
an attractor diagram. So far only a small part of this diagram has
been explored experimentally by Metzger et al. with the photothermal effect as the 
driving force \cite{metzger2008self}. Little effort has been made to investigate the attractor 
diagram experimentally using radiation pressure force. It has therefore been to date an open
problem to explore the full attractor diagram experimentally \cite{aspelmeyer2013cavity}.
One reason for this is that a transition from self-induced oscillations to chaotic mirror motion
can occur due to second order effects such as absorption-induced heating of the optical
components \cite{carmon2005temporal,marino2013coexisting}.
This restricts the exploration of the attractor diagram to small values of the mirror amplitude.

Here we demonstrate an optomechanical setup,
consisting of a Fabry-Perot cavity with a trampoline resonator,
that does not suffer from optical absorption in the mirrors. Not only does this enables us to explore a large part of the attractor diagram in a controlled fashion, we also find surprisingly rich dynamics and non-trivial steady states of our optomechanical system.

\section{Theoretical model}

Our optomechanical system is described by two coupled equations
of motion: 
\begin{eqnarray}
\frac{d\alpha(t)}{dt} & = & \frac{-\kappa}{2}\alpha(t)+i(\Delta+Gx(t))\alpha(t)+\sqrt{\kappa_{ex}}\alpha_{in}\\
\frac{d^2x(t)}{dt^2} & = & -\Omega_{m}^{2}x(t)-\Gamma_{m}\frac{dx(t)}{dt}+\frac{\hbar G}{m}\left|\alpha(t)\right|^{2}
\end{eqnarray}
in which $\alpha$ is the cavity field and $x$ the mirror displacement.
The parameters in Eqs. 1-2 are defined as follows: 
$\alpha_{in}$ is the laser field, $\kappa$ the
overall cavity decay rate, $\kappa_{ex}$ the cavity entrance loss
rate, $\Delta=\omega_{L}-\omega_{cav}$ the laser detuning defined
as the difference between cavity and laser frequency, the optical
frequency shift per displacement $G=\omega_{cav}/L$, with $L$ being
the length of the cavity, $\Omega_{m}$
the fundamental mode frequency of the mechanical oscillator, $\Gamma_{m}$
the mechanical damping rate and $m$ the mode mass of the harmonic
oscillator. Thermal and mechanical noise sources have been neglected; an important
assumption that will be justified for our optomechanical system by the results below.

The optomechanical attractor diagram displays the optomechanical gain
 $\zeta_{opt}$, the ratio of the radiative force $P_{rad}$ and frictional losses $P_{fric}$, as a function of laser detuning $\Delta$ and mirror amplitude $A$. 
From Eqs. 1-2 an expression for  $\zeta_{opt}$ can be derived \cite{marquardt2006dynamical}:
\begin{equation}
\zeta_{opt}(\Delta,A)=\frac{P_{rad}}{P_{fric}}=-\frac{1}{\Gamma_{m}}\frac{2\hbar G\kappa_{ex}\alpha_{in}^{2}}{m\Omega_{m}A}\textrm{Im}(\sum_{n}\alpha_{n+1}^{*}\alpha_{n})
\end{equation}
with 
\begin{equation}
\alpha_{n}=\frac{J_{n}(-GA/\Omega_{m})}{\kappa/2-i\tilde{\Delta}+in\Omega_{m}}
\end{equation}
in which $\alpha_{n}$ is the nth harmonic (or sideband)
in the optical field created by the mirror motion, $J_{n}$ the Bessel function of the first kind and $\tilde{\Delta}$
the effective laser detuning defined as $\tilde{\Delta}=\omega_{L}-\omega_{cav}+G\bar{x}$
where $\bar{x}$ is the static displacement of the mirror due to the
radiation pressure. For most situations, including ours, the static displacement is
negligible and $\Delta\thickapprox\tilde{\Delta}.$ 
Stable self-induced oscillations occur when $\zeta_{opt}(\Delta,A)=1$, while amplification (attenuation) 
of the mechanical mode occurs when $\zeta_{opt}(\Delta,A)>1$ ($\zeta_{opt}(\Delta,A)<0$).
For our system the Brownian motion at 300K is already sufficient to achieve $\zeta_{opt}(\Delta,A)>1$.

One way to map out the attractor diagram $\zeta_{opt}(\Delta,A)$ is to 
measure the mirror amplitude while varying the laser detuning.
Such measurement schemes have already successfully been used for demonstrating optical cooling. 
With optical cooling, the change in cavity resonance frequency due to the motion of the mirror is usually much smaller than the 
linewidth of the cavity resonance, i.e. $GA\ll\kappa$. In the optical field only the first sideband is visible and the magnitude of this sideband is linear with mirror amplitude. For optical excitation, however, the change in cavity resonance frequency can be much larger than the cavity linewidth, i.e. $GA\gg\kappa$, resulting in multiple sidebands present in the optical field. The linear relation between the first sideband and mirror amplitude no longer holds. Now the mirror amplitude can only be obtained by taking into account all optical sidebands.

\section{Experimental setup}

\begin{figure}
\centering{}\includegraphics[scale=0.3]{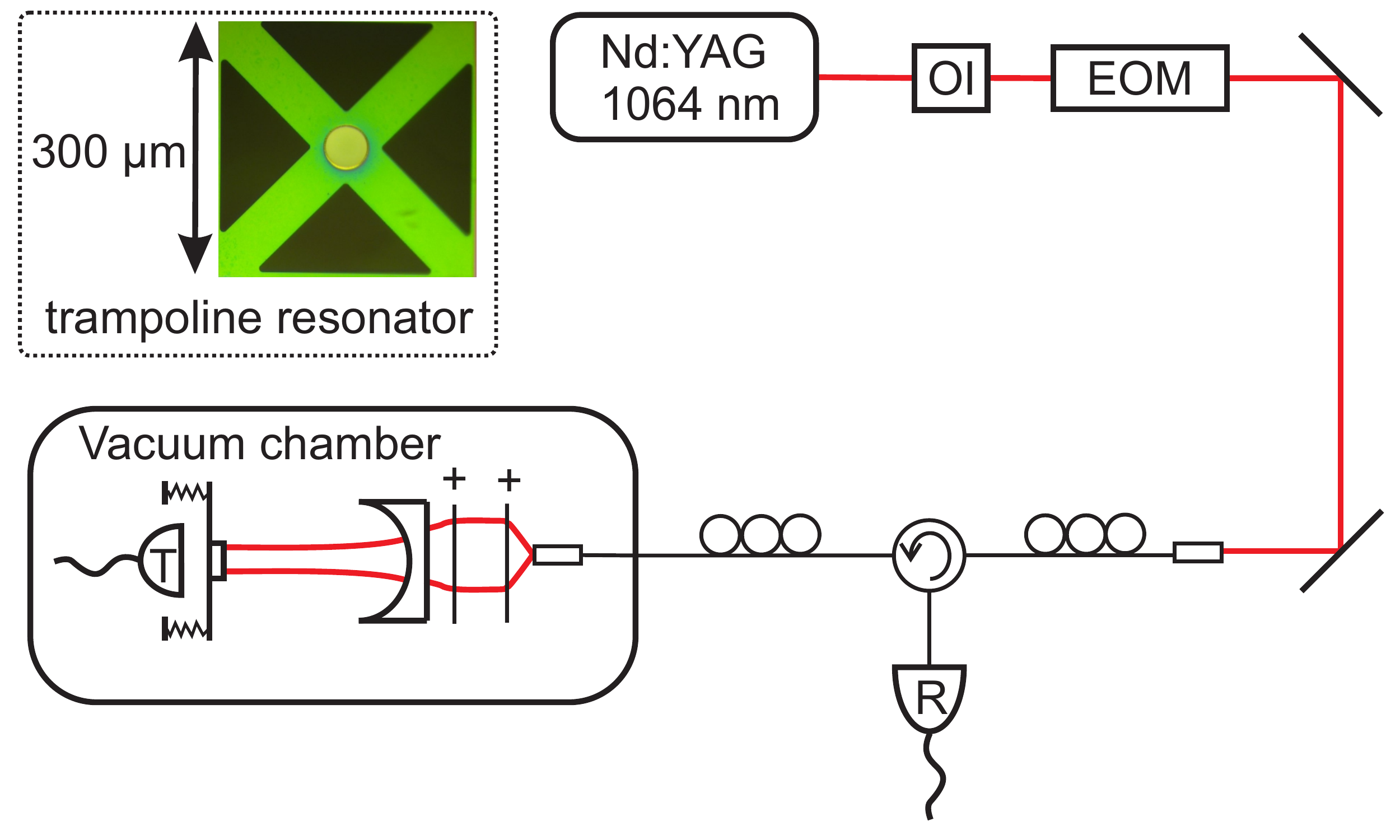}\caption{Schematic overview of 
the set up. A piezo tunable CW Nd:YAG laser
is passed through an optical isolator (OI) and a 9.5 MHz electro-optical modulator (EOM) before it
enters a fiber circulator that is fed into a vacuum chamber that contains a 5 cm Fabry-Perot cavity. Both
the transmitted and the reflected intensity are recorded with photo-detectors.
The inset shows an optical image of the trampoline resonator.}
\end{figure}

To map out the attractor diagram we make use of a 5 cm long Fabry-Perot cavity
operating around 1064 nm, with a trampoline resonator as one of the
end mirrors \cite{kleckner2011optomechanical}. By using a multilayer Bragg stack on both cavity mirrors, absorption losses
are minimized to about 1 ppm.
The system is placed inside a vacuum chamber with a vibration isolation system containing
several Eddy-current dampers. All measurements are performed at room temperature.
A schematic overview of the set-up is
given in Fig. 1. We use a piezo tunable CW Nd:YAG laser and control
it with a typical scan speed of $\frac{d\omega_{L}}{dt}=100 - 400$ MHz/s, which is slow compared
to the cavity build up time, i.e. $\frac{d\omega_{L}}{dt}\ll\kappa/\tau$
with $\kappa$ the cavity linewidth and $\tau$ the cavity lifetime.
An electro-optical modulator (EOM)
at 9.5 MHz is used to calibrate the laser detuning.
The mechanical properties of the trampoline resonator are characterized
by measuring the thermal noise spectrum and the optical properties
by performing an optical ring-down measurement \cite{kleckner2011optomechanical}.
Both transmitted and reflected cavity light are detected using photo-detectors
and the data-acquisition is done using a digital storage scope. For
our system only the fundamental mechanical mode and fundamental optical mode (TEM$_{00}$)
are relevant. The parameters for our system are the following: $\kappa=175\times10^{3}\times2\pi$
rad/s, $\kappa_{ex}=50\times10^{3}\times2\pi$ rad/s, $\Omega_{m}=343\times10^{3}\times2\pi$
rad/s, $\Gamma_{m}=1.7\times2\pi$ rad/s at a pressure of $10^{-6}$ mbar and $m=110\times10^{-12}$
kg. To achieve a sufficiently large optomechanical gain, the input laser power should also be sufficiently large.
A typical laser input power of 50 to 100 \textrm{$\mu$}W is used, corresponding to an intracavity photon number of $2.8-5.6 \times 10^8$.

\section{Results}
\begin{figure}
\centering{}\includegraphics[scale=0.37]{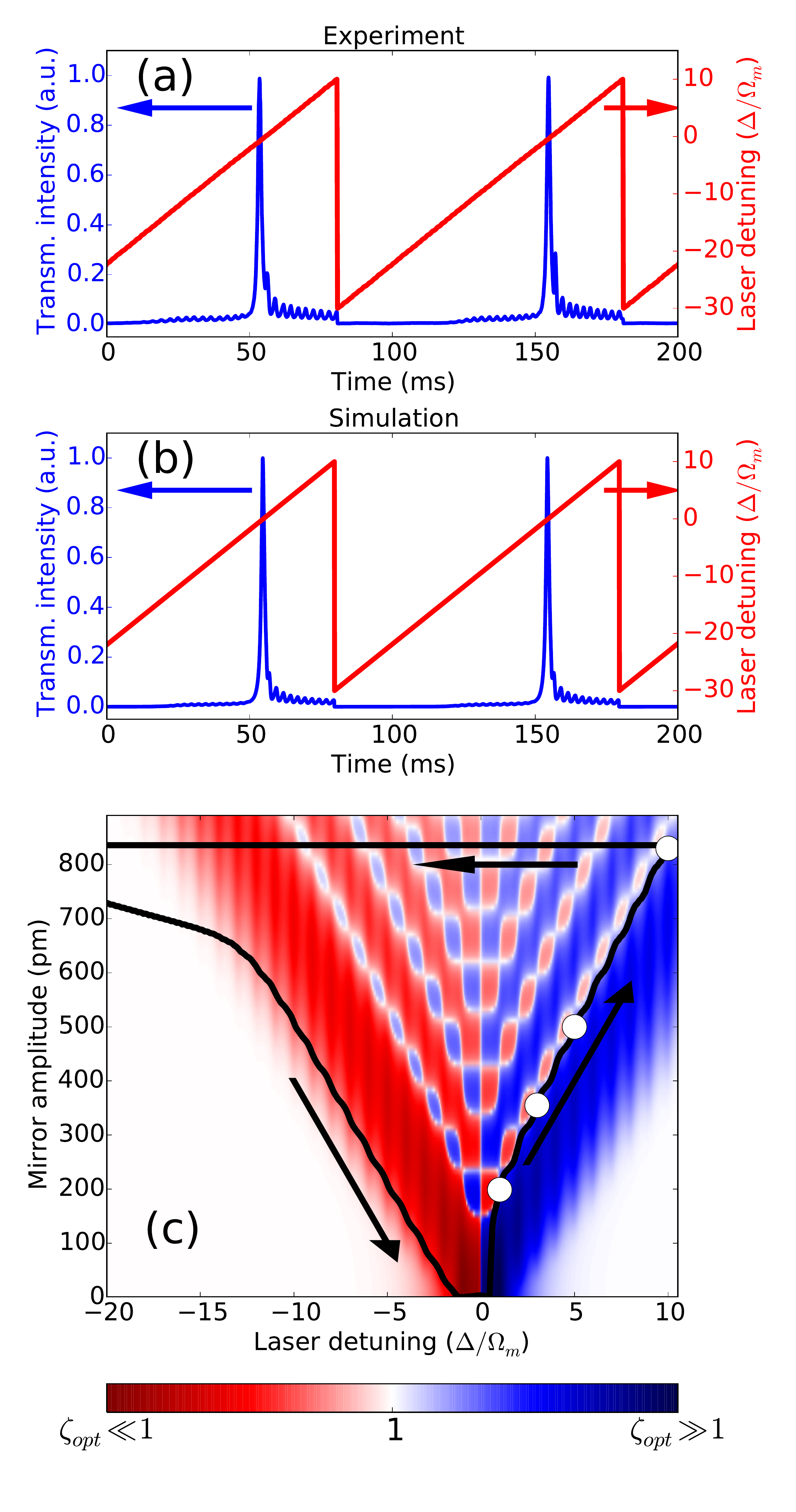}\caption{(color online) A closed cycle across the attractor diagram. (a) Intensity transmitted by the cavity for two consecutive periods of a controlled laser detuning sweep.  (b) Simulation based on Eqs. 1-2.  (c) Attractor diagram corresponding to our experimental parameters. The path followed in the experiment is indicated by the arrows.}
\end{figure}

Fig. 2a shows the optical intensity transmitted 
by the cavity when the laser is scanned back and forth across the cavity resonance. 
Several peaks are visible, not only at the cavity resonance $\Delta/\Omega_{m}=0$ but also at 
multiples of $\Delta/\Omega_{m}$. The appearance of sidebands can be explained as follows.
Suppose the laser frequency is at $\omega_{L}=\omega_{cav}+\Omega_{m}$
and the amplitude of the mirror is small such that only the first
sideband is created by the moving mirror at frequencies $\omega=\omega_{L}\pm\Omega_{m}$.
Only the Stokes sideband at $\omega=\omega_{cav}$ is resonant with
the cavity and enhanced, while the anti-Stokes sideband at $\omega=\omega_{cav}+2\Omega_{m}$
is suppressed. So the interaction of the blue detuned laser field
with the resonator leads to a resonant field in the cavity. The non-linear
interaction of the resonant cavity field plus the incoming laser field
with the mirror lead to a resonant driving force. By creating sidebands,
the mirror generates its own driving force, which increases the mirror
amplitude. The increased mirror amplitude leads to a stronger modulation
of the cavity field, and this process repeats until the gain is reduced to
$\zeta_{opt}=1$ (limit cycle behavior). When the laser detuning is
slowly increased further, the process repeats whereby the ever increasing
mechanical motion allows sideband generation to drive the mirror to
larger amplitudes.
This process continues until the laser is swept back rapidly to  $\Delta/\Omega_{m}=-30$. 
At first the laser detuning and mirror amplitude do not match to produce an optical 
force that influences the mirror motion. The mirror amplitude decreases only due to 
the intrinsic mechanical damping. While the laser detuning is slowly increased towards zero detuning, 
at some point the laser detuning and mirror amplitude are such that sidebands created by the 
mirror motion result in an optical force. However the sign of the optical force has changed compared 
to the situation with positive detuning. Instead of parametric amplification, now parametric attenuation 
occurs, resulting in a decrease in mirror amplitude. The interaction of the laser field with the resonator again leads to a resonant cavity field, resulting in peaks at multiples of $\Delta/\Omega_{m}$ also for negative laser detunings. This is only visible when the mirror amplitude was driven to large values previously. Driven oscillations at negative laser detunings reveal therefore something about the state and history of the system and are non-trivial.

To compare the experimental result of Fig. 2a with theory, a numerical simulation is performed 
with the same experimental parameters. For this we solve numerically Eqs. 1-2 using the following initial conditions: 
$\alpha(0)=0$, $\alpha'(0)=0$, $x(0)=x_{0}$ and $x'(0)=0$
where $x_{0}$ denotes the initial mirror amplitude $x_{0}$. The value for $x_{0}$ is chosen to correspond to the thermal mirror amplitude at 300K: $x_{0}=\sqrt{k_{b}T/m\Omega^2_{m}}$. Although no mechanical and thermal noise is required to reproduce the experimental results, an initial  mirror amplitude is needed to start the parametric process.

The results of the simulation, depicted in Fig. 2b, are in good agreement with the experimental results of Fig. 2a. This indicates that our earlier assumption not to include thermal and mechanical noise in Eqs. 1-2 is justified. Furthermore, we do not need to include any second order effects such as heating of the mirror substrates due to absorption. Although from the experimental data the mirror amplitude is not obtained directly, the numerical simulations do contain the mirror amplitude. By plotting the attractor diagram according to Eq. 4 together with the mirror amplitude obtained from the simulations, we can visualize the traversed path across the attractor diagram.

In Fig. 2c the attractor diagram is displayed together with the evolution of the mirror amplitude (indicated by the arrows). 
The amplitude follows a deterministic path through the diagram. Along this path the optomechanical gain varies. When the gain is large, the path closely follows the $\zeta_{opt}=1$ contour, while in the regions with moderate gain the changing laser detuning prevents the mirror amplitude from reaching the $\zeta_{opt}=1$ contour as closely. Specifically, for positive laser detunings  $\zeta_{opt}\geq1$ and for negative laser detunings $\zeta_{opt}\leq1$. It is also worthwhile to emphasize that the mirror amplitude changes on the time scale of the laser scan speed, much slower than the oscillation frequency of the mirror or the cavity lifetime. So far we have thus been discussing the dynamics of a driven, quasi-static, system. However, also  interesting dynamics occur on the time scale of the mechanical resonator. 

\begin{figure}
\centering{}\includegraphics[scale=0.33]{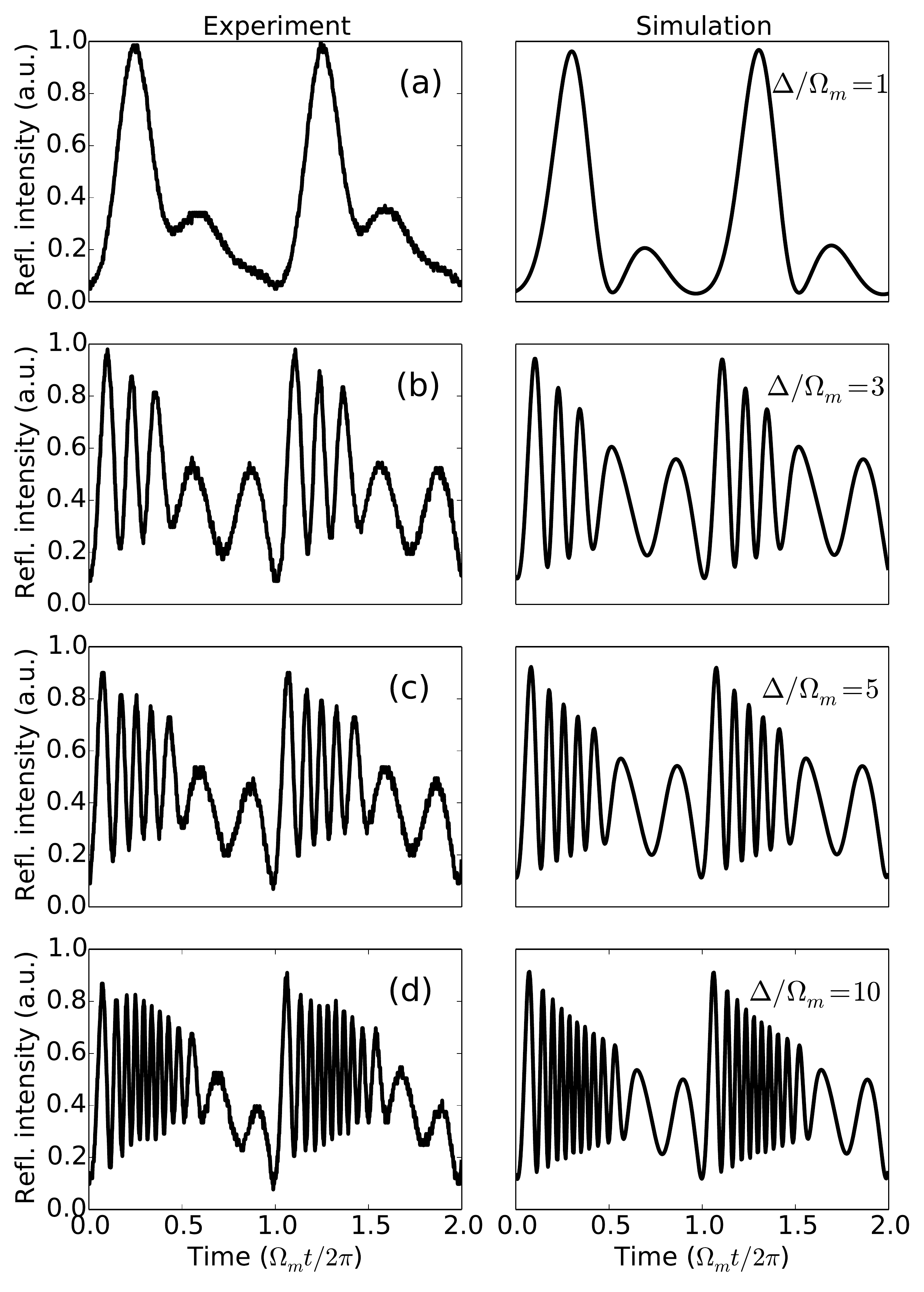}\caption{Detailed time traces of the reflected cavity intensity for different
laser detunings. The left column shows the measurements and the right column numerical solutions to
Eqs. 1-2. (a)-(d) correspond to specific detunings indicated with white dots in Fig. 2c.}
\end{figure}

Theoretically the increase of the mirror amplitude, as shown in Fig. 2c,  should be visible as an increase in the number of harmonics present
in the output field \cite{marquardt2006dynamical}. This is verified by analyzing the fast modulation present in the reflected intensity for several different detunings corresponding to the white dots in Fig. 2c. We have analyzed the reflected intensity as
it is picked up by a faster photodetector in our experimental configuration. However the same features should also be visible in the transmitted intensity.

In Fig. 3 we compare experimental and numerical results for these fast modulations. For clarity an offset is removed and the figures rescaled. The excellent agreement between theory and experiment confirms once more that we have explored in detail the boundary (lowest branch where $\zeta_{opt}=1$) of the attractor diagram and that this method is suited for further exploration of the attractor diagram. Furthermore, we have significantly amplified the motion of our mechanical resonator, using large intracavity power, without any sign of unstable or chaotic behavior. 

\begin{figure}
\centering{}\includegraphics[scale=0.37]{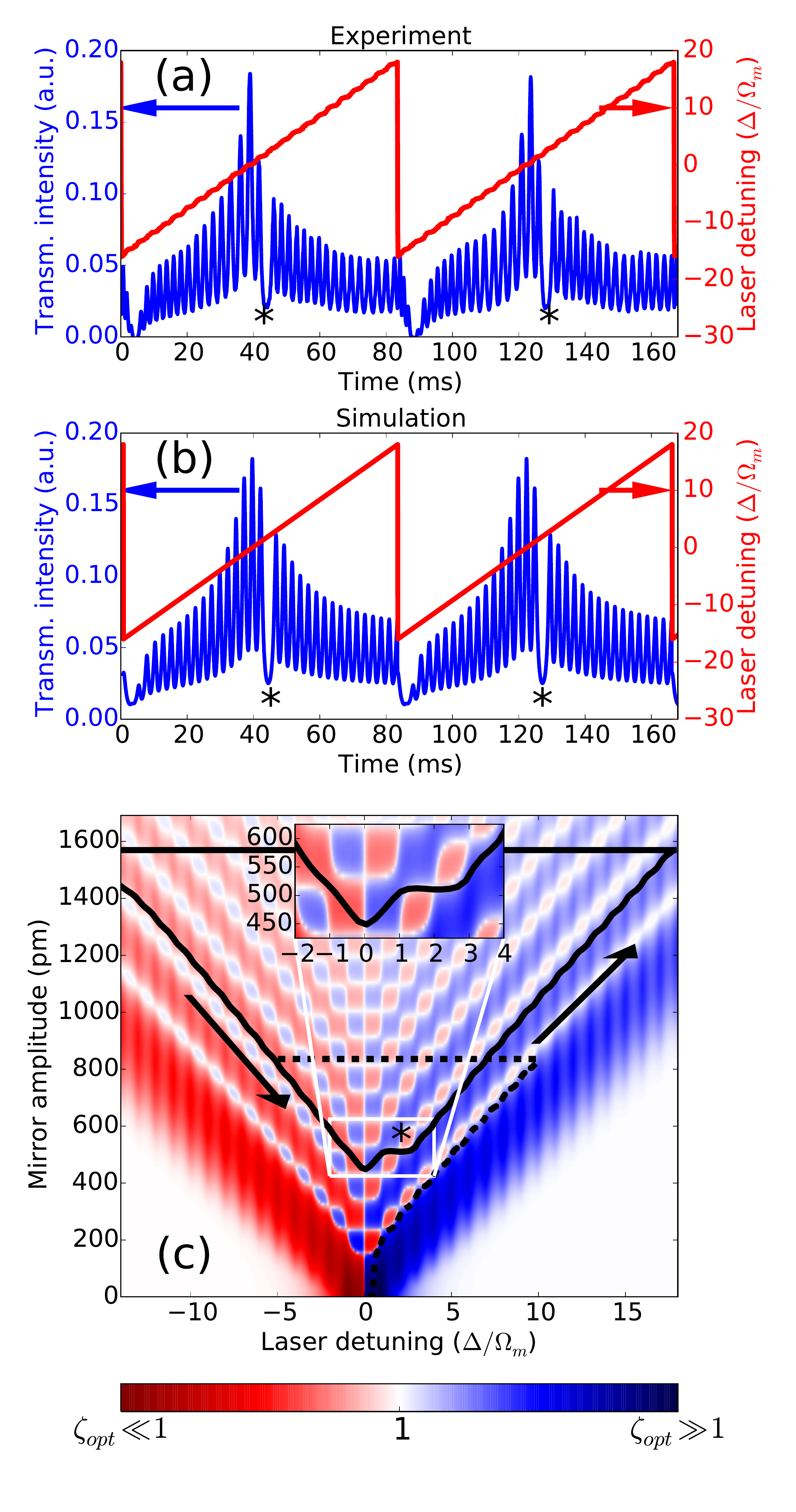}\caption{(color online) Exploring a higher branch in the attractor diagram. (a) Intensity transmitted by the cavity for two consecutive periods of a controlled laser detuning sweep. The scale for the transmitted intensity is the same as in Fig. 2a (b) Simulation based on Eqs. 1-2. (c) Attractor diagram together  with the path followed in the experiment. Before switching to a higher branch, the system is initialized (dashed line) using a similar detuning sweep as in Fig. 2c. Inset: Zoom of region around zero detuning.}
\end{figure}
 
To demonstrate that we have full control over our system, we change the starting conditions of our laser frequency sweep after performing a cycle similar to the one displayed in Fig. 2. When the mirror amplitude is large, changing the laser detuning slightly makes it possible to skip from the boundary branch to another branch. In this way different branches in the attractor diagram can be explored.

Fig. 4a shows the results of two cycles across the attractor diagram along a different branch. The scale for the transmitted intensity is the same as in Fig. 2a. Although the experimental conditions have only changed a little, the result is quite different from Fig. 2a. Still multiple peaks at $\Delta/\Omega_{m}$ are visible, but the main cavity resonance at $\Delta/\Omega_{m}=0$ is reduced significantly compared to these sidebands. Also a distinctive dip is visible, indicated with "*". To verify that the features of Fig. 4a are captured by the theoretical model of Eqs. 1-2, a numerical simulation is performed with the same experimental parameters. The qualitative agreement between experiment and simulation shows that the model is still valid for our system. Furthermore, from the simulation we can again extract the mirror amplitude and use this together with the attractor diagram to explain the features of Fig. 4a.

Fig. 4c shows this attractor diagram. The black dashed line shows the initialization, which is similar to the cycle performed in Fig. 2, but now the laser detuning is set back to just $\Delta/\Omega_{m}=-5$ to reach a different branch. Note that the initialization is not shown in Fig. 4a and 4b. The solid black line shows the evolution of the mirror amplitude during one cycle. The largest mirror amplitude achieved in this experiment is roughly 1600 pm, more than 500 times the amplitude at 300K without any sign of chaotic or unstable behavior.

For the steady state cycles of Fig. 4a the reduction of the transmitted intensity at the cavity resonance ($\Delta/\Omega_{m} = 0$) is now readily explained: the large mirror amplitude reduces the time the cavity is resonant with the input field, therefore less intracavity field is built-up, resulting in a reduction of transmitted intensity.

The inset of Fig. 4c shows the region where a change from one stable branch to another occurs. This transition occurs at $ \left\{ \Delta/\Omega_{m}=1.5, A\approx 510\,\mathrm{pm}\right\}$. At this point the mirror amplitude stays constant along a contour where $\zeta_{opt} = 1$. This point coincides with the distinctive dip in Fig. 4. When the mirror amplitude does not change, no optical driving force occurs and no sideband is visible in the optical output. Even more interesting is the surrounding area of the attractor diagram. At $ \left\{ \Delta/\Omega_{m}=1.5, A\approx 510\,\mathrm{pm}\right\}$ any small change in mirror amplitude is significantly amplified: if the mirror amplitude increases slightly, $\zeta_{opt} \gg 1$ and the mirror amplitude will increase significantly. Similarly, if the mirror amplitude decreases slightly, $\zeta_{opt} \ll 1$ and the mirror amplitude will decrease significantly. The inset therefore highlights a bistability: a small perturbation of the mirror motion will result in a large change in the mirror amplitude. However, our results show that in a clean system such as ours, we can "walk" through such unstable regions on a well-defined path.

\section{Conclusion}
With the absence of any chaotic or unstable behavior our optomechanical system is only described by two equations (Eqs. 1-2). This has allowed us to explore in detail the optomechanical attractor diagram and the dynamics of self-induced oscillations. By performing a laser frequency sweep, multiple stable branches in the attractor diagram are explored. Furthermore, we have demonstrated non-trivial dynamics such as driven oscillations for negative laser detunings and the presence of a bistability.

\begin{acknowledgments}
The authors would like to thank H. van der Meer for technical assistance
and support. The authors are also grateful for useful discussions
with M.P. van Exter, W. Loeffler and G. Welker. 
This work is part of the research program of the Foundation for Fundamental Research on Matter (FOM) and of the NWO VICI research program, which are both part of the Netherlands Organisation for Scientific Research (NWO). This work is also supported by National Science Foundation Grant No. PHY-1212483.

\end{acknowledgments}

\end{justify}
\end{document}